\documentclass[aps,prb,amsmath,amssymb,reprint]{revtex4-1}

\usepackage{amsmath}
\usepackage{amssymb}
\usepackage{graphicx}
\usepackage{dcolumn}
\usepackage{bm}
\usepackage{hyperref}

\usepackage[abs]{overpic}
\usepackage{pict2e}
\usepackage{bbm}

\renewcommand{\Im}{{\mathrm {Im}}}

\newcommand{\grad}{\nabla}
\renewcommand{\part}{\partial}

\newcommand{\rom}[1]{\textup{\uppercase\expandafter{\romannumeral#1}}}

\newcommand{\Bh}{{\bf h}}

\newcommand{\Bj}{{\bf j}}
\newcommand{\Bk}{{\bf k}}

\newcommand{\Bm}{{\bf m}}

\newcommand{\Bp}{{\bf p}}

\newcommand{\Br}{{\bf r}}

\newcommand{\Bv}{{\bf v}}

\newcommand{\BB}{{\bf B}}

\newcommand{\BE}{{\bf E}}
\newcommand{\BF}{{\bf F}}

\newcommand{\BM}{{\bf M}}

\newcommand{\BP}{{\bf P}}
\newcommand{\BQ}{{\bf Q}}
\newcommand{\BR}{{\bf R}}

\numberwithin{equation}{section}

\begin{document}

\preprint{APS/123-QED}

\title{Nonlinearity in the Lorentz Oscillator Model}

\author{Brad C. Smith}
\email{bradcs@umich.edu}
\affiliation{Applied Physics, University of Michigan, Ann Arbor, MI, USA}

\begin{abstract}
The magnetic force is retained in the Lorentz Oscillator Model and a perturbation solution is derived beyond the dipole approximation. Perturbation series for the electric dipole, magnetic dipole, and electric quadrupole moments in addition to their macroscopic, spatially-averaged counterparts are also obtained. These expressions are shown to describe many optical phenomena: the normal, linear Zeeman, Faraday, and magneto-optic Kerr effects, the linear Stark, Pockels, and Kerr effects (when the restoring force is anharmonic), the momentum and angular momentum of a photon in matter (contributing to the Abraham-Minkowski debate), the photon drag and inverse Faraday effects, and sum-frequency and second-harmonic generation in centrosymmetric media. The qualitative and quantitative accuracy is assessed by comparison to both experiment and more fundamental theory (e.g.\ quantum mechanics). New insights into the relationship between the Stark and Pockels/Kerr effects also emerge.
\end{abstract}

\maketitle

\section{\label{sec:intro}Introduction}

The Lorentz Oscillator Model (LOM) has been a staple in optics for over 100 years \cite{Lorentz1909}. It's original formulation played a major role in the acceptance of the electron due to its explanation of the Zeeman effect \cite{Lorentz1897,Zeeman1896,Zeeman1896b,Kox1997}. The calculated electron charge to mass ratio agreed with J. J. Thomson's experiment the following year which further confirmed the particle's existence \cite{Thomson1897first,Buchwald2001}. Today, the LOM is useful for gaining an intuitive understanding of the physical interaction between light and matter and is frequently seen in an educational context. When only the electric term is retained in the Lorentz force and its spatial dependence ignored (electric dipole approximation) the LOM captures a surprising number of features (dispersion, absorption, etc.) for such a simple model. The expression obtained for the atomic polarizability even has the same structure as one obtained via quantum mechanics and also satisfies Kramers-Kronig relations \cite{Zangwill2013}.

In its original conception, the LOM utilized the full Lorentz force \cite{Lorentz1897}. But as it became supplanted by quantum mechanics in practice, the LOM's primary function gradually became more pedagogical. In this role, the magnetic force term is frequently dropped to focus on the physical effects of the electric field \cite{Milonni1988}. In this work, the magnetic term is kept and the name LOM used here refers to a model that includes both the electric and magnetic forces. Although Lorentz retained the magnetic force from an applied static magnetic field, the force from the optical magnetic field was still neglected \cite{Lorentz1897}. In modern textbooks it is argued that since the magnetic force from a plane wave is about $v/c$ (where $v$ is a particle's speed and $c$ is the speed of light) times smaller in magnitude than the corresponding electric force it may be neglected for driving fields of non-relativistic strength \cite{Milonni1988}. However, these forces are usually orthogonal and occur at different frequencies. So, while it may be small, the optical magnetic force is sometimes the only force that drives motion in a particular direction or at a particular frequency. Because of this the contributions of the magnetic force can be significant even for low intensities of light. For example, the optical magnetic force is responsible for the transfer of light's momentum to matter and leads to the photon drag effect (discussed in Sec.\ \ref{sec:pde}).


Several works have included the optical magnetic force but only considered special cases within the electric dipole approximation: in Ref.\ \cite{Fisher2010} the electric and magnetic fields are restricted to be monochromatic and non-static while Ref.\ \cite{Oughstun2006} cannot support a static magnetic field. In this work the full magnetic force is retained in addition to anharmonic restoring forces and perturbation solutions are obtained beyond the electric dipole approximation. These solutions qualitatively and quantitatively (usually within an order of magnitude) describe magneto-optic effects (the normal, linear Zeeman, Faraday, and surface magneto-optic Kerr effects), electro-optic effects (the linear Stark, Pockels, and Kerr effects), the momentum and angular momentum of a photon in matter (contributing to the Abraham-Minkowski debate), and other three-wave mixing processes allowed in centrosymmetric media (the photon drag and inverse Faraday effects and sum-frequency and second-harmonic generation). Although this extended version of the LOM may not capture the finer details found using semi-classical or fully quantized approaches, its simplicity makes it an ideal choice for building intuition, crude calculations, and qualitative study.


\section{\label{sec:eoms}Equations of motion}

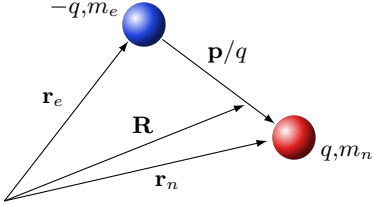
\begin{figure}[ht]
\centering
\hspace*{\fill}
\begin{picture}(50,28)

 
 \put(15,20) {\includegraphics[scale=0.01]{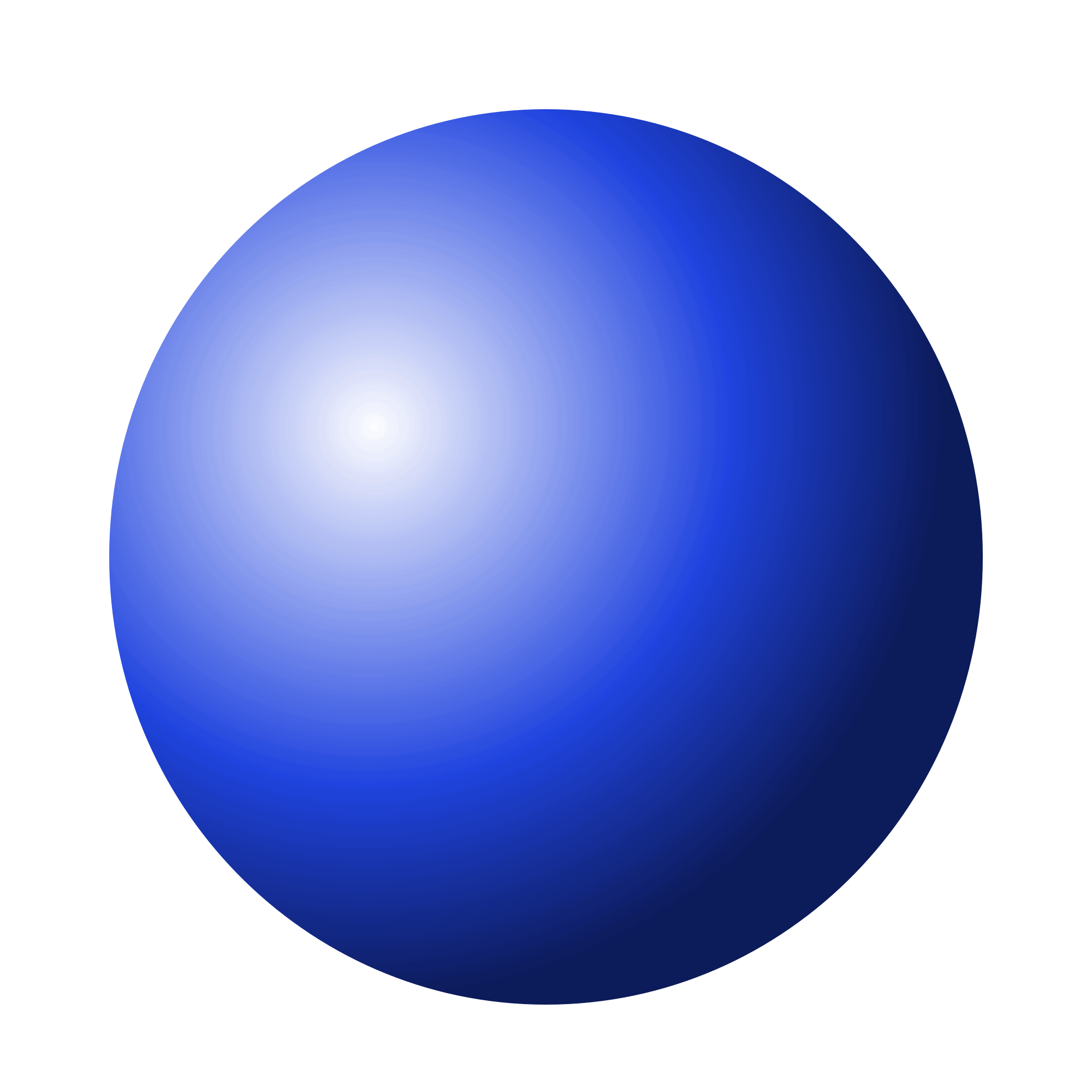}}
 \put(6,25) {$-q$,$m_e$}
 \put(0,0){\vector(15.5,20){16.5}}
 \put(5,13) {$\Br_e$}
 
 \put(35,5) {\includegraphics[scale=0.01]{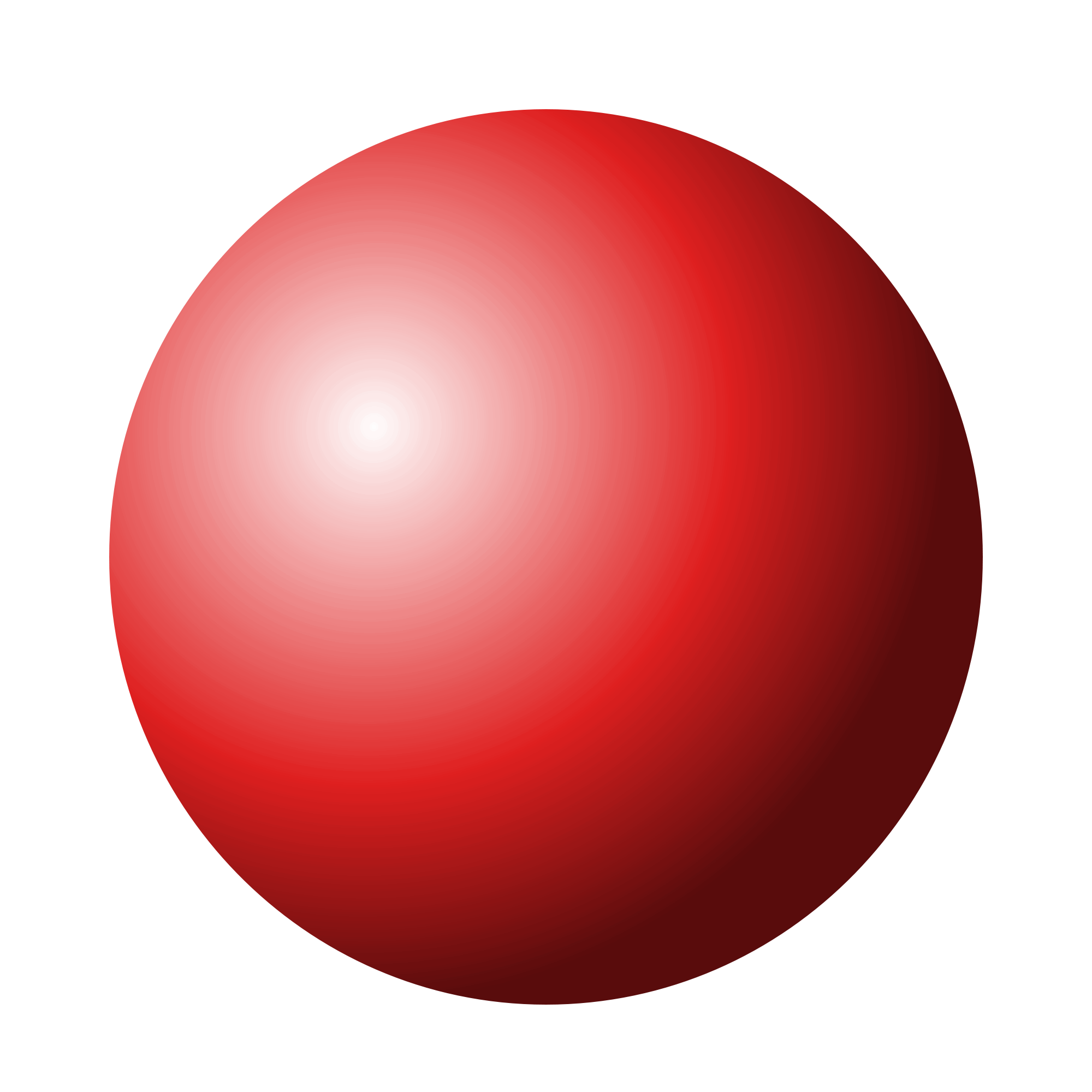}}
 \put(42,6) {$q$,$m_n$}
 \put(0,0){\vector(35,8){35}}
 \put(20,2) {$\Br_n$}
 
 \put(21,21.5){\vector(20,-15){15}}
 \put(27,19) {$\Bp/q$}
 
 \put(0,0){\vector(30,12){32}}
 \put(17,9) {$\BR$}
 
\end{picture}
\hspace*{\fill}

\caption{Coordinates in the LOM.}
\label{fig:lom}
\end{figure}

A description of the LOM written by H.\ A.\ Lorentz can be found in Ref.\ \cite{Lorentz1909}. It was known at the time that a charged particle oscillating at a certain frequency would radiate light having that same frequency. From this, Lorentz conjectured that the spectral lines in a material could be the result of harmonically bound pairs of oppositely charged particles with each spectral line corresponding to a different binding resonant frequency. A modern description which more closely follows this work can be found in Ref.\ \cite{Milonni1988}. The LOM consists of two point masses referred to here as the nucleus, with mass $m_n$ and charge $q$, and electron, with mass $m_e$ and charge $-q$ (see Fig.\ \ref{fig:lom}). With no external forces, the equilibrium positions of each particle are assumed to be the same as the result of a conservative restoring force between the two particles. For small excursions from equilibrium a Taylor series expansion of the corresponding potential energy yields a Hooke's law type force as the leading term followed by anharmonic terms. Damping forces are inserted phenomenologically to account for energy loss (see Ref.\ \cite{Siegman1986} for a detailed discussion). Lastly, the electric and magnetic fields are coupled to each particle via the Lorentz force.

\subsection{\label{sec:eomsB}Newton's second law}

The equations of motion for the positions of the electron, $\Br_e(t)$, and nucleus, $\Br_n(t)$ are provided by Newton's second and third laws:
\begin{align} 
m_n \ddot{\Br}_n &= \BF_{Ln} + \BF_{dn} + \BF_r  \label{eq:eomn} \\
m_e \ddot{\Br}_e &= \BF_{Le} + \BF_{de} - \BF_r\; . \label{eq:eome}
\end{align}
The Lorentz forces are $\BF_{Ln} = q\,\BE(\Br_n) + q\,\dot{\Br}_n \times \BB(\Br_n)$ and $\BF_{Le} = -q\,\BE(\Br_e) - q\,\dot{\Br}_e \times \BB(\Br_e)$ where $\BE$ and $\BB$ are the local electric and magnetic fields respectively, the phenomenological damping forces are $\BF_{dn} = -m_n\boldsymbol{\gamma}\cdot\dot{\Br}_n$ and $\BF_{de} = -m_e\boldsymbol{\gamma}\cdot\dot{\Br}_e$ where $\boldsymbol{\gamma}$ is a symmetric rank-two tensor, and the restoring force is $\BF_r = -\Bk^{(1)} \cdot \Bp / q- \Bk^{(2)} : \Bp\Bp / q^2 - \Bk^{(3)} \; \scalebox{0.7}{\vdots} \; \Bp\Bp\Bp / q^3 + \cdots$ where $\Bp$ is the electric dipole moment defined in Eq.\ \eqref{eq:defp} and $\Bk^{(1)}$, $\Bk^{(2)}$, and $\Bk^{(3)}$ are symmetric tensors of rank two, three, and four, describing the first-order (harmonic), second-order (anharmonic), and third-order (anharmonic) restoring forces respectively (see Ref.\ \cite{Boyd2008}).

It is convenient to express the equations of motion in the center of mass coordinate system defined by
\begin{align}
\BR &\equiv \frac{m_e\Br_e + m_n\Br_n}{m_e + m_n} &  &\text{and} & \Bp &\equiv q\left( \Br_n - \Br_e \right)  \label{eq:defp}\big.
\end{align}
where $\BR(t)$ is the position of the center of mass of the system and $\Bp(t)$ is the electric dipole moment (see Fig.\ \ref{fig:lom}). Using this coordinate system and the explicit forms of the forces, the equations of motion are
\begin{align}
\ddot{\BR} + \boldsymbol{\gamma}\cdot\dot{\BR} &= \frac{1}{M}\Bigg\{ q\bigg[\BE(\Br_n) - \BE(\Br_e) \bigg]    \nonumber \\
&+ q\,\dot{\BR} \times \bigg[\BB(\Br_n) - \BB(\Br_e) \bigg] \nonumber \\
&+ \frac{\dot{\Bp}}{M} \times \bigg[ m_e \, \BB(\Br_n) + m_n\, \BB(\Br_e) \bigg] \Bigg\}  \label{eq:eomR}
\end{align}
for the center of mass and
\begin{align}
\ddot{\Bp} + \boldsymbol{\gamma}\cdot\dot{\Bp} + \frac{\Bk^{(1)}}{m}\cdot\Bp &= \frac{q}{m}\Bigg\{ qm\bigg[\frac{\BE(\Br_n)}{m_n} + \frac{\BE(\Br_e)}{m_e} \bigg]    \nonumber \\
&+ qm\,\dot{\BR} \times \bigg[\frac{\BB(\Br_n)}{m_n} + \frac{\BB(\Br_e)}{m_e} \bigg] \nonumber \\
&+ \frac{m\, \dot{\Bp}}{M} \times \bigg[ \frac{m_e \, \BB(\Br_n)}{m_n} - \frac{m_n\, \BB(\Br_e)}{m_e} \bigg] \nonumber \\
&- \frac{\Bk^{(2)}}{q^2} : \Bp\Bp - \frac{\Bk^{(3)}}{q^3} \;\vdots\; \Bp\Bp\Bp \Bigg\} \label{eq:eomp}
\end{align}
for the electric dipole moment where $M\equiv m_n + m_e$ and $m\equiv m_nm_e/M$ are the system's total and reduced mass respectively.

\subsection{\label{sec:eomsC}Taylor expansion}

Eqs. \eqref{eq:eomR} and \eqref{eq:eomp} require knowledge of $\BE$ and $\BB$ at the positions of each particle for all time making it difficult to find solutions directly. As long as the maximum displacement of the particles from $\BR$ is much less than the wavelength of the fields it is justifiable to expand the electric and magnetic fields in a Taylor series about the center of mass position $\BR$ \cite{Milonni1988}. Retaining only the first term corresponds to making the usual electric dipole approximation. The following terms incorporate higher-order multipole moments. Each series will be truncated such that the equations of motion are valid to second order in the perturbation expansion presented in Sec.\ \ref{sec:sol}. Thus, up to second-order the equations of motion become
\begin{align} \label{eq:eomtemp1}
\ddot{\BR} + \boldsymbol{\gamma}\cdot\dot{\BR} &= \frac{1}{M}\bigg[ \left(\Bp\cdot\grad \right)\BE + \dot{\Bp} \times \BB \bigg] \; ,
\end{align}
for the center of mass and
\begin{align} \label{eq:eomtemp2}
\ddot{\Bp}+\boldsymbol{\gamma}\cdot\dot{\Bp} +\frac{\Bk^{(1)}}{m}\cdot\Bp &= \frac{q}{m}\bigg[ q\,\BE -\xi_2\left(\Bp\cdot\grad\right)\BE  \nonumber \\
&+ q\,\dot{\BR} \times \BB - \xi_2\,\dot{\Bp} \times \BB \nonumber \\
&- \frac{\Bk^{(2)}}{q^2} : \Bp \Bp \bigg] \; ,
\end{align}
for the electric dipole moment where the gradients and all fields are evaluated at $\BR$ and where we have defined the dimensionless factor $\xi_2 \equiv (m_n^2 - m_e^2)/M^2$ which ranges from $-1$ to $1$. See Ref.\ \cite{Zangwill2013} for a detailed discussion of the force terms in Eq.\ \eqref{eq:eomtemp1}.

\section{\label{sec:sol}Perturbation expansion}

Contrary to the typical linear response LOM these equations of motion are nonlinear and have no known analytical solutions. For this reason, perturbation theory can be used to obtain perturbation solutions \cite{Boyd2008}. For simplicity $\Bv \equiv \dot{\BR}$ is the velocity of the atom. First, $\Bv$ and $\Bp$ are expanded in a power series of the perturbation parameter $\lambda$
\begin{align}
\Bv &= \Bv^{(0)} + \lambda\,\Bv^{(1)} + \lambda^2\,\Bv^{(2)} + \cdots \label{eq:expandv}  \\   
\Bp &= \Bp^{(0)} + \lambda\,\Bp^{(1)} + \lambda^2\,\Bp^{(2)} + \cdots \; . \label{eq:expand}
\end{align}
Note that $\Bv^{(0)}$ and $\Bp^{(0)}$ represent homogeneous solutions (which are ignored in this work) and all other terms represent particular solutions. Next, the driving fields are treated as perturbations by taking $\BE \rightarrow \lambda \BE$ and $\BB \rightarrow \lambda \BB$. Inserting these expressions into Eqs.\ \eqref{eq:eomtemp1} and \eqref{eq:eomtemp2} yields two equations in many powers of $\lambda$. When the coefficients of each power of $\lambda$ are gathered and set equal to zero, two infinite series of simultaneous equations are produced. The first few equations of these series are
\begin{align} 
\dot{\Bv}^{(1)} + \boldsymbol{\gamma}\cdot\Bv^{(1)} &= \boldsymbol{0} \; , \label{eq:v1}
\end{align}
\begin{align} 
\dot{\Bv}^{(2)} + \boldsymbol{\gamma}\cdot\Bv^{(2)} &= \frac{1}{M}\bigg[ \left(\Bp^{(1)}\cdot\grad \right)\BE + \dot{\Bp}^{(1)} \times \BB \bigg] \; , \label{eq:v2}
\end{align}
for the velocity and
\begin{align}
\ddot{\Bp}^{(1)}+\boldsymbol{\gamma}\cdot\dot{\Bp}^{(1)} +\frac{\Bk^{(1)}}{m}\cdot\Bp^{(1)} &= \frac{q^2}{m}\,\BE \; , \label{eq:p1}
\end{align}
\begin{align} 
\ddot{\Bp}^{(2)}+\boldsymbol{\gamma}\cdot\dot{\Bp}^{(2)} +\frac{\Bk^{(1)}}{m}\cdot\Bp^{(2)} =& \frac{q}{m}\bigg[ q\,\Bv^{(1)} \times \BB \nonumber \\
&-\xi_2\left(\Bp^{(1)}\cdot\grad\right)\BE \nonumber \\
&- \xi_2\,\dot{\Bp}^{(1)} \times \BB \nonumber \\
&- \frac{\Bk^{(2)}}{q^2} : \Bp^{(1)} \Bp^{(1)} \bigg] \; . \label{eq:p2}
\end{align}
for the electric dipole moment.

Solutions to each equation can be obtained in the frequency domain using Fourier transforms of the form:
\begin{align}
\Bh(t) &= \int\limits_{-\infty}^\infty \! \mathrm{d}\omega \; \tilde{\Bh}(\omega)\;e^{-i\omega t} \; .
\end{align}
For example, the solution to the traditional linear LOM (see Eq.\ \eqref{eq:p1}) is $\Bp^{(1)}(\omega)=\varepsilon_0\,\boldsymbol{\alpha}^{(1)}(\omega)\cdot\tilde{\BE}(\omega)$ where the linear polarizability is
\begin{align}
\boldsymbol{\alpha}^{(1)}(\omega) &\equiv \frac{q^2}{m\varepsilon_0} \left[\frac{\Bk^{(1)}}{m} - \omega^2\boldsymbol{\delta} - i\omega \boldsymbol{\gamma} \right]^{-1} \; ,
\end{align}
and $\boldsymbol{\delta}$ is the unit dyadic. The solution for $\Bp^{(2)}$ was omitted here for brevity, but can be obtained by dividing Eq.\ \eqref{eq:P2} by $N_s$ (see Sec.\ \ref{sec:const} for more information).

\subsection{\label{sec:solA}Other multipole moments}

Perturbation solutions can also be obtained for other multipole moments of the oscillator as well. For example, the magnetic dipole moment in the center of mass frame (which is the appropriate frame to use later for spatial averaging - see Ref.\ \cite{Jackson1998}) is:
\begin{align} \label{eq:magmoment}
\Bm &\equiv \frac{1}{2}\sum\limits_\alpha q_\alpha (\Br_\alpha - \BR) \times (\dot{\Br}_\alpha - \dot{\BR}) \nonumber \\ 
&= -\frac{\xi_2}{2q} \Bp \times \dot{\Bp} \; .
\end{align}
Since our solution for $\Bp$ is already a perturbation expansion in the fields, a perturbation solution for $\Bm$ can be obtained by expanding
\begin{align}
\Bm &= \Bm^{(0)} + \lambda\,\Bm^{(1)} + \lambda^2\,\Bm^{(2)} + \cdots \; .
\end{align}
with the same perturbation parameter $\lambda$ and inserting Eq.\ \eqref{eq:expand} into Eq.\ \eqref{eq:magmoment}. Gathering terms of equal power in $\lambda$ reveals
\begin{align}
\Bm^{(1)} &= \boldsymbol{0} \; , \label{eq:m1}
\end{align}
\begin{align}
\Bm^{(2)} &= -\frac{\xi_2}{2q} \,\Bp^{(1)} \times \dot{\Bp}^{(1)} \; . \label{eq:m2}
\end{align}

In a similar manner, perturbation solutions may be obtained for the electric quadrupole moment
\begin{align}
\boldsymbol{\mathbbmss{q}}^{(1)} &= \boldsymbol{0} \; , \label{eq:q1}
\end{align}
\begin{align}
\boldsymbol{\mathbbmss{q}}^{(2)} &= -\frac{\xi_2}{2q} \,\Bp^{(1)} \Bp^{(1)} \;. \label{eq:q2}
\end{align}
The magnetic quadrupole and electric octupole moments appear at third-order.

\section{\label{sec:const}Constitutive relations}

The LOM can be used to obtain macroscopic constitutive relations for the polarization density, $\BP$, the magnetization density, $\BM$, and the electric quadrupolarization density, $\BQ$ (and higher order terms if desired) by treating matter as a collection of non-interacting oscillators each with their own resonant and damping frequencies and following the spatial averaging procedure in Ref.\ \cite{Jackson1998}. Assuming non-interacting oscillators is analogous to treating matter as an ensemble of non-interacting two-level, quantum mechanical systems -- each oscillator represents a single two-level system. However, real materials often exhibit responses that imply coupling between more than two energy levels (and thus coupling between oscillators) which is beyond the scope of this work. Despite this, the extension of the LOM presented here can still describe many real optical phenomena.


For simplicity local field corrections are ignored (i.e.\ the local fields in Sec.\ \ref{sec:sol} are assumed equal to the macroscopic fields). Furthermore, the constitutive relations below are written assuming a homogeneous material consisting only of a single species of oscillator with number density $N_s$ (in this case, $\BP(\Br,t)$ is simply $N_s \, \Bp(t)$ where the fields in $\Bp$ are evaluated at $\Br$). Wherever it is necessary to consider multiple species the expressions below can be summed or integrated over all species. For example, in resonant effects (e.g.\ the photon drag effect - see Sec.\ \ref{sec:pde}) a large response can come entirely from a single species of oscillator that may comprise a very small subset of the medium.


\subsection{\label{sec:constA}Polarization density}

Since the solution for $\Bp$ is already a perturbation expansion in $\BE$ and $\BB$, a perturbation solution for $\BP$ can be obtained by expanding
\begin{align}
\BP &= \BP^{(0)} + \lambda\,\BP^{(1)} + \lambda^2\,\BP^{(2)} + \cdots \; . \label{eq:Pexpand}
\end{align}
with the same perturbation parameter $\lambda$ and following the spatial averaging procedure of Ref.\ \cite{Jackson1998}. Gathering terms of equal power in $\lambda$ up to second-order and expressing the results in the frequency-domain reveals
\begin{align}
\tilde{\BP}_i^{(1)}(\omega) &= \varepsilon_0 N_s \,\boldsymbol{\alpha}^{(1)} (\omega) \cdot \tilde{\BE} (\omega) \; , \label{eq:P1}
\end{align}
and
\begin{align}
\tilde{P}_i^{(2)}(\omega) &= \varepsilon_0  \int\limits_{-\infty}^\infty \! \mathrm{d}\omega_1 \int\limits_{-\infty}^\infty \! \mathrm{d}\omega_2 \;\delta(\omega_1 + \omega_2 - \omega) \nonumber \\
&\quad \Bigg\{ N_s\,\alpha_{ijkl}^{(L)} (\omega_1,\omega_2)  \bigg[i\omega_1\tilde{E}_j(\omega_1)\epsilon_{klm}\tilde{B}_m(\omega_2) \nonumber \\
&\quad + \tilde{E}_j(\omega_1) \grad_k \tilde{E}_l(\omega_2) \bigg] \nonumber \\
&\quad + N_s\,\alpha_{ijk}^{(2)}(\omega_1,\omega_2)\, \tilde{E}_j(\omega_1)\tilde{E}_k(\omega_2) \Bigg\} \; , \label{eq:P2}
\end{align}
where
\begin{align}
\alpha_{ijkl}^{(L)}(\omega_1,\omega_2) &\equiv \frac{-\xi_2\varepsilon_0}{q}\, \alpha_{il}^{(1)}(\omega_1 + \omega_2)\,\alpha_{kj}^{(1)}(\omega_1)  \; ,  \label{eq:chipeb}
\end{align}
which is responsible for the photon drag, magneto-optic Kerr, Faraday, and linear Zeeman effects, and
\begin{align}
\alpha_{ijk}^{(2)}(\omega_1,\omega_2) &\equiv \frac{-\varepsilon_0^2}{q^3} \, k_{lmn}^{(2)}\, \alpha_{il}^{(1)}(\omega_1 + \omega_2)\,\alpha_{mj}^{(1)}(\omega_1) \,\alpha_{nk}^{(1)}(\omega_2) \; , \label{eq:chi2}
\end{align}
which is responsible for the Pockels and linear Stark effects.

\subsection{\label{sec:constB}Magnetization}

In a similar manner to $\BP$, the perturbation solution for $\BM$ up to second-order is
\begin{align}
\BM^{(1)} &= \boldsymbol{0} \; , \\
\BM^{(2)} &= -\frac{\xi_2 N_s}{2q} \Bp^{(1)}\times\dot{\Bp}^{(1)} \; . \label{eq:M2alt}
\end{align}
In the frequency-domain the last equation may be written as
\begin{align}
\tilde{M}_i^{(2)}(\omega) &= \epsilon_0  \int\limits_{-\infty}^\infty \! \mathrm{d}\omega_1 \! \int\limits_{-\infty}^\infty \! \mathrm{d}\omega_2 \; \delta(\omega_1 + \omega_2 - \omega) \nonumber \\
&N_s \, \alpha_{jklm}^{(M)}(\omega_1,\omega_2) \tilde{E}_j(\omega_1)\,\epsilon_{ikl} \, i\omega_2\tilde{E}_m(\omega_2) \; , \label{eq:M2}
\end{align}
where
\begin{align} \label{eq:IFEsus}
\alpha_{jklm}^{(M)}(\omega_1,\omega_2) & \equiv -\frac{\xi_2\varepsilon_0}{2q} \alpha_{lj}^{(1)}(\omega_1)\,\alpha_{km}^{(1)}(\omega_2) \; ,
\end{align}
which is responsible for the Inverse Faraday effect. Eqs.\ \eqref{eq:M2alt}-\eqref{eq:IFEsus} agree with other classical calculations based on the Lorentz force \cite{Hertel2006,Battiato2014}.

\subsection{\label{sec:constC}Electric quadrupolarization}

Likewise the perturbation solution for $\BQ$ up to second-order is
\begin{align}
\BQ^{(1)} &= \boldsymbol{0} \; , \\
\BQ^{(2)} &= -\frac{\xi_2 N_s}{2q} \Bp^{(1)} \Bp^{(1)} \; .
\end{align}
In the frequency-domain the last equation may be written as
\begin{align}
\tilde{Q}_{ij}^{(2)}(\omega) &= \varepsilon_0  \int\limits_{-\infty}^\infty \! \mathrm{d}\omega_1 \! \int\limits_{-\infty}^\infty \! \mathrm{d}\omega_2 \; \delta(\omega_1 + \omega_2 - \omega) \nonumber \\
&N_s \, \alpha_{ijkl}^{(Q)}(\omega_1,\omega_2) \tilde{E}_k(\omega_1)\tilde{E}_l(\omega_2) \; . \label{eq:Q2}
\end{align}
where 
\begin{align} 
\alpha_{ijkl}^{(Q)}(\omega_1,\omega_2) & \equiv -\frac{\xi_2\varepsilon_0}{2q} \, \alpha_{ik}^{(1)}(\omega_1)\,\alpha_{jl}^{(1)}(\omega_2) \; . 
\end{align}
All of the susceptibilities defined above possess intrinsic permutation symmetry as expected \cite{Boyd2008}.

\section{\label{sec:effects}Physical phenomena}

Maxwell's equations require constitutive relations in order to be complete - without them it is impossible to predict the evolution of $\BE$ of $\BB$ in the presence of matter. Furthermore, the constitutive relations physically describe the interaction between the electromagnetic fields and matter and therefore contain all of the physical effects observed in experiments. The standard undergraduate approach (ignore $\BB$, $\BM$, $\BQ$, and anharmonicity) captures a surprising number of linear optical effects. But by extending the LOM in the manner above many more optical effects can be accounted for both qualitatively and roughly quantitatively.

Every polarizability in the perturbation solutions for $\BP$, $\BM$, and $\BQ$ (see Eqs. $\eqref{eq:P1}$, $\eqref{eq:P2}$, $\eqref{eq:M2}$, and $\eqref{eq:Q2}$) gives rise to one or more physical effects. The second-order effects are collectively known as three-wave mixing processes and are explored below.

\subsection{\label{sec:mo}Linear magneto-optic effects}

Linear magneto-optic effects occur when the optical properties of a material have a linear dependence on a quasi-static magnetic field. Clearly these effects arise from the first term in Eq.\ \eqref{eq:P2}. When a static magnetic field with strength $\BB_0$ is present in an isotropic material the second-order response that depends on this field (given by Eq.\ \eqref{eq:P2}) is
\begin{align}
\tilde{\BP}^{(FZ)}(\omega) &= i\omega\varepsilon_0\,N_s\alpha^{(L)} (\omega,0)\; \BB_0\times\tilde{\BE}(\omega)  \; . \label{eq:FZP}
\end{align}
When this polarization is inserted into the wave equation the cross product necessarily implies that its contribution to the susceptibility is strictly off-diagonal and anti-symmetric which is the hallmark of the Faraday, Zeeman, and magneto-optic Kerr effects (MOKE) \cite{Lorentz1897,Yariv1984,Qiu2000}. The qualitative and quantitative descriptions of the Faraday and linear, ``normal'' Zeeman effects provided by the LOM are quite remarkable (see \ref{sec:fzdetails}). However, the MOKE and its primary application - the surface magneto-optic Kerr effect (SMOKE) - generally use magnetic materials because the change in polarization is much larger. Since the LOM neglects both spin and exchange interactions which are fundamental to magnetism it cannot accurately predict the size of these effects in magnetic materials. Therefore, the main practical value in the LOM for describing the MOKE and SMOKE is that it correctly attributes these effects to consequences of the magnetic component of the Lorentz force.

\subsection{\label{sec:eo}Linear electro-optic effects}

Linear electro-optic effects occur when the optical properties of a material have a linear dependence on a quasi-static electric field. These effects arise from the anharmonic (last) term in Eq.\ \eqref{eq:P2}. When a static electric field with strength $\BE_0$ is present in a non-centrosymmetric material the second-order response that depends linearly on this field (given by Eq.\ \eqref{eq:P2}) is
\begin{align}
\tilde{\BP}^{(SP)}(\omega) &= 2\varepsilon_0\,N_s\boldsymbol{\alpha}^{(2)} (0,\omega) : \BE_0\tilde{\BE}(\omega)  \; . \label{eq:SPP}
\end{align}
When this polarization is inserted into the wave equation the effective susceptibility is modified leading to the Pockels effect \cite{Yariv1984} (and linear Stark effect as argued below). The qualitative agreement with experimental results on the Pockels effect is excellent while the quantitative agreement (when using a reasonable estimation for $\Bk^{(2)}$) is less admirable but still decent \cite{Boyd2008}.

In \ref{sec:fzdetails} the connection between the Faraday and linear Zeeman effects as a reflection of the Kramers-Kronig relations was discussed. Since the Pockels and linear Stark effects are the electric analogs of the Farday and linear Zeeman effects respectively, one might suspect that they too are different aspects of the same physical interaction. In fact, W.\ Voigt's predictions for the Stark effect in 1901 were based on adding an anharmonic restoring force term to the LOM \cite{Voigt1901,Kox2013}. Modern literature generally only associates the Pockels effect with the anharmonic restoring force making the linear Stark effect appear unrelated. However, there is simply no other way to represent a small linear resonance shift in a constitutive relation besides through $\boldsymbol{\alpha}^{(2)}$ (or $\boldsymbol{\chi}^{(2)}$). To remedy this, a simple argument for presenting both effects as $\boldsymbol{\alpha}^{(2)}$ effects is presented in \ref{sec:SP}.

In principle linear electro-optic effects could also arise from Eqs.\ \eqref{eq:M2}, \eqref{eq:Q2}, and the middle term of Eq.\ \eqref{eq:P2}. However, in unstructured media their magnitudes are estimated to be at least $d/\lambda$ times smaller than the standard $\boldsymbol{\chi}^{(2)}$ effects for off-resonant excitation where $d$ is the lattice constant and $\lambda$ is the wavelength of light. This comes by comparing the magnitudes of the polarizabilities in Eqs.\ \eqref{eq:chi2} and \eqref{eq:chipeb} and using the standard estimate of $k^{(2)} = m\omega_0^2/d$ \cite{Boyd2008}. Still, in a centrosymmetric material (where $\boldsymbol{\chi}^{(2)}=\boldsymbol{0}$) it might be possible to observe them. On the other hand these effects could become appreciable in nano-scale devices or meta-materials where field gradients much stronger than $E/\lambda$ can exist.

\subsection{\label{sec:dfgrect}Difference-frequency generation and optical rectification}

Difference-frequency generation (DFG) and optical rectification (OR) occur when a material responds at the difference between two (typically optical) field frequencies. All of the second-order constitutive relations (Eqs.\ \eqref{eq:P2}, \eqref{eq:M2}, and \eqref{eq:Q2}) support DFG and OR. The standard $\chi^{(2)}$ effects are covered extensively in the literature (e.g.\ Ref.\ \cite{Boyd2008}) so only the remaining terms will be discussed here.

The difference frequency part of the first term in Eq.\ \eqref{eq:P2} is a consequence of the transfer of linear momentum between the optical fields and matter known as the photon drag effect (PDE) - it is discussed in Sec.\ \ref{sec:pde}. Eq.\ \eqref{eq:M2} leads to the Inverse Faraday effect which is the rotational analog of the PDE - it is discussed in Sec.\ \ref{sec:ife}. The quadrupole response in Eq.\ \eqref{eq:Q2} is of similar magnitude and is largest for two non-collinear beams or a single tightly-focused one. The second term of Eq.\ \eqref{eq:P2} combined with the PDE term has the same structure as the force on the center-of-mass (see Eq.\ \eqref{eq:v2}) that gives rise to optical tweezers \cite{Zangwill2013}. Therefore, the LOM predicts that optical tweezers should polarize objects as they are accelerated so long as $m_e \neq m_n$. This, and the other effects in this section, lead to the conclusion that light preferentially imparts its momentum, angular momentum, and energy, to the lighter charge carrier which may subsequently transfer these quantities to its heavier partner via the restoring force.

\subsubsection{\label{sec:pde}Photon drag effect}

One of the key features of the LOM is the ability to microscopically track the flow of energy and momentum. This makes the LOM a natural choice for describing the photon drag effect (PDE) which is characterized by the transfer of the Minkowski momentum ($n'\hbar \omega / c$) per quantum of light to the absorbing electrons in a material where $n'$ is the real part of the refractive index of the medium and $\omega$ is the angular frequency of the light \cite{Gibson1970,Danishevski1970,Gibson1980}. This momentum transfer polarizes the material parallel or anti-parallel to the direction of propagation of light. The PDE arises from the time-averaged (or difference-frequency) part of the first term of Eq. $\eqref{eq:P2}$. However, since the PDE is most often seen in semiconductors one may first convert the LOM to the Drude model by taking $\omega_0 \rightarrow 0$ and examining the time-averaged current $\langle\Bj_P\rangle \equiv \langle\dot{\BP}\rangle$.

\ref{sec:photmom} tracks the energy and momentum of light quanta inside a material by comparing the power and force provided by a plane wave propagating in a material. To summarize the results, the momentum of each quantum of light with energy $\hbar\omega_0$ propagating inside a material is the average of the Abraham and Minkowski values (given by Eq.\ \eqref{eq:totphotmom}) which is exclusively transferred to the electrons (assuming $m_e\ll m_n$). An electron resonantly absorbing light acquires the Minkowski momentum per photon (Eq.\ \eqref{eq:minkow}) while the remaining electrons receive the remaining momentum. These results agree with those derived using quantum optical and other classical methods \cite{Mansuripur2004,Loudon2005}. By considering pulses of light the momentum transfer at surfaces versus in bulk can be distinguished \cite{Mansuripur2005,Loudon2005}. 

The voltage predicted by the LOM for typical PDE experiments (open-circuit conditions) can be obtained by first calculating the electric field that counteracts the time-averaged Lorentz force acting on the resonant absorbers. Turning Eq.\ \eqref{eq:forcedense2} into a force density acting on the electrons of identical oscillators under resonant illumination yields
\begin{align}
\left\langle \BF_{Le}^{(1,2)} \right \rangle &= \frac{\kappa I_0}{ c} e^{-\kappa z}\,\hat{\Bk} \; , \label{eq:eleforce}
\end{align}
where $I_0$ is the light's peak intensity (Poynting vector magnitude) in vacuum ignoring reflections, $\kappa$ is the extinction coefficient, and the light is normally incident on the material's surface lying at $z=0$. This assumes $N_s\alpha'' = \chi''$ where $N_s$ is the number density of resonant absorbers (free carriers). This division of the medium into resonant oscillators (which only contribute to $\chi''$) and host oscillators (which only contribute to $\chi'$) is appropriate for the materials used \cite{Birch1974}. Eq.\ \eqref{eq:eleforce} is identical to the first equation in Ref.\ \cite{Gibson1980} - the induced voltage can be calculated by following that work which agrees very well with their experimental results. If all of the light is absorbed, an order of magnitude estimate of the voltage is $V\approx I_0/(qcN_s)$ where $N_s$ is the number density of resonant absorbers. Recall that $I_0/c$ is also the radiation pressure from light. To summarize, optical momentum transfer can polarize a material and/or drive a current so long as $m_e \neq m_n$.


\subsubsection{\label{sec:ife}Inverse Faraday effect}

The Inverse Faraday effect (IFE) is essentially the angular momentum analog of the photon drag effect. The extension of the LOM presented here predicts that circularly polarized photons each transfer an angular momentum of magnitude $\hbar$ to the electrons in a material. This transfer of angular momentum magnetizes the medium parallel or anti-parallel to the direction of propagation of light. The IFE arises from the time-averaged (or difference-frequency) part of Eq. $\eqref{eq:M2}$.

\ref{sec:photangmom} tracks the energy and angular momentum of light quanta inside a material by comparing the power and torque provided by a plane wave propagating in a material. To summarize the results, the angular momentum of each quantum of light with energy $\hbar\omega_0$ propagating inside a material (given by Eq.\ \eqref{eq:photangmom}) is $\boldsymbol{0}$ for linearly polarized light and $\hbar \hat{\Bk}$ $(-\hbar \hat{\Bk})$ for left (right) circularly polarized light from the viewpoint of the receiver which is exclusively transferred to the electrons (assuming $m_e\ll m_n$). Unlike the linear momentum, this angular momentum is independent of the refractive index and the photon's energy and all of it is transferred to the electron that absorbs the photon.

The magnetic field created by the magnetization in Eq.\ \eqref{eq:M2} must satisfy Maxwell's equations and can be found using the Biot-Savart law. Note that although plane waves can produce a magnezation, they cannot induce a magnetic field -- this would violate Gauss's law for magnetism and is a reflection of the fact that plane waves carry zero total angular momentum \cite{Sokolov1991}. So, Gaussian beams will be considered instead. For a medium composed of many species of oscillator, two limiting cases for the induced magnetic field are presented here - both ignore reflections and assume a monochromatic Gaussian beam with waist $w_0$ (the radius where the intensity falls to $1/e^2$ of its peak value) and Jones vector $\boldsymbol{\mathcal{E}}$ is normally incident on a slab of material. Note that any material with a large Faraday effect should also have a large IFE \cite{Pershan1966}.

Off resonance the magnetic field induced in the middle of a material whose length is much greater than $w_0$ is
\begin{align}
\left\langle \BB \right\rangle &\approx \frac{i\pi\xi_2  I_0 }{q\lambda_0 N c^2} \frac{(n'^2 - 1)^2}{n'^2} \,\hat{\boldsymbol{\mathcal{E}}} \times \hat{\boldsymbol{\mathcal{E}}}^* \; ,
\end{align}
where $N$ is the number density of all oscillators. This expression assumes the magnitudes of each polarizability are roughly equivalent far from any resonance. An order-of-magnitude estimate for the magnetic field induced by visible light in typical condensed matter is $1$ $\mu$G/MW/cm\textsuperscript{2} which agrees well with experiment \cite{Ziel1965}. In magnetic materials (which are not described by the LOM) the effect can be orders of magnitude greater \cite{Kimel2005}.

On resonance the magnetization is a manifestation of the optical orientation of spin carriers that is of interest to the spintronics community \cite{Zutic2004}. According to the LOM the induced magnetic field at the surface in the middle of the beam is
\begin{align}
\left\langle \BB \right\rangle &= \frac{i\xi_2 I_v \lambda_0 \kappa^2}{8\pi q c^2N_s} \, f\!\left(\frac{w_0\kappa}{2}\right)  \; \hat{\boldsymbol{\mathcal{E}}} \times \hat{\boldsymbol{\mathcal{E}}}^* \; ,
\end{align}
where $I_0$ is the light's peak intensity (Poynting vector magnitude) in vacuum ignoring reflections, $\kappa$ is the extinction coefficient, and
\begin{align}
f\left(x\right) &\equiv 1 - \sqrt{\pi} \, x + x^2 e^{-x^2} \bigg[\pi f_i(x) - f_e(x^2) \bigg] \; ,
\end{align}
where $f_i(x) \equiv \text{erf}(ix)/i$ is the imaginary error function and $f_e(x) \equiv -\int_{-x}^\infty e^{-x}/x \; \mathrm{d}x$ is the exponential integral function. This expression uses the same approximation as Sec.\ \ref{sec:pde} namely that $N_s\alpha'' = \chi''$ where $N_s$ is the number density of resonant absorbers. It also assumes \emph{all} of the light is resonant. For a sense of scale, if all of the nuclear spins in GaAs point in the same direction the electrons see a magnetic field of about 5 T \cite{Paget1977}. This would correspond to an electron spin polarization of about 1\% which can easily be achieved by exciting a material with resonant circularly polarized light \cite{Zutic2004}. Similar to the PDE, the IFE is a consequence of the conservation of angular momentum. The transfer of angular momentum from optical fields to the charge carriers in a material will magnetize it so long as $m_e \neq m_n$.





\subsection{\label{sec:sfgshg}Sum-frequency and second-harmonic generation}

Sum-frequency generation (SFG) and second-harmonic generation (SHG) occur when a material responds at the sum of two (typically optical) field frequencies. All of the second-order constitutive relations (Eqs.\ \eqref{eq:P2}, \eqref{eq:M2}, and \eqref{eq:Q2}) support SFG and/or SHG. The standard $\chi^{(2)}$ effects are covered extensively in the literature (e.g.\ Ref.\ \cite{Boyd2008}) so only the remaining terms will be discussed here.

As discussed in Sec.\ \ref{sec:eo} the other second-order polarizabilities are expected to be at least approximately $d / \lambda$ times smaller than $\boldsymbol{\alpha}^{(2)}$ where $d$ is the lattice constant and $\lambda$ is the wavelength of light. This limits the practical value of these other terms. Still, they have been observed in natural, centrosymmetric media \cite{Bethune1976,Bethune1981,Terhune1962}. It is worth noting these effects are largest when two orthogonally-polarized, non-collinear excitation beams are used. Furthermore, Eq.\ \eqref{eq:M2} implicitly forbids second-harmonic magnetization in isotropic media.


\section{\label{sec:concl}Conclusion}

This work extends the LOM by retaining $\BB$ in the equations of motion and also deriving expressions for the multipole moments $\Bm$ and $\boldsymbol{\mathbbmss{q}}$. The physical phenomena described via this extension are now united in a single classical model rather than phenomenological or quantum mechanical ones. Given this, the accuracy (both qualitative and quantitative) of the LOM is remarkable (discussed in Sec.\ \ref{sec:effects}). The advantage of this presentation is the simplicity of working with a two-body system and Newton's second law. This is ideal for educational settings, order-of-magnitude calculations, and building physical intuition.

There are several limitations to this presentation of the LOM worth noting. First, since spin is neglected it cannot describe magnetic materials, the anomalous Zeeman effect, etc. Second, there is no inclusion of acoustic (e.g.\ Raman or Brillouin) effects. Third, there are several ``free'' parameters (e.g.\ resonance frequency or oscillator strength) that must be known a priori (usually by experiment or quantum-mechanical calculation). Ultimately the trade-off made with the LOM is the surrender of veracity for simplicity which is indeed desirable under certain circumstances.


\section{\label{sec:ack}Acknowledgements}

The author acknowledges illuminating discussions with Albert Liu, Steven Cundiff, Duncan Steel, and Herbert Winful at the University of Michigan.

\appendix
\numberwithin{equation}{section}

\section{\label{sec:fzdetails}Faraday and Zeeman effects}

Since Eq.\ \eqref{eq:FZP} can be cast in a form similar to Eq.\ \eqref{eq:P1} one can define an effective linear polarizability of the oscillator as
\begin{align}
\alpha_{ij}(\omega) &= \alpha^{(1)}(\omega)\,\delta_{ij} + i\omega \,\alpha^{(L)} (\omega,0)\,B_{0,k}\, \epsilon_{ikj} \; .
\end{align}
The Levi-civita symbol in the second term implies that its contribution is strictly anti-symmetric and off-diagonal. For a macroscopic medium composed of this single species of oscillators this leads to an induced circular birefringence in isotropic materials \cite{Yariv1984}. For plane waves propagating parallel to the applied magnetic field the eigenmodes are left and right circularly polarized with corresponding susceptibilities (via Eq.\ \eqref{eq:chipeb}) given by
\begin{align}
\chi_{\pm}(\omega) &= \chi^{(1)}(\omega) \pm  \frac{\xi_2\varepsilon_0\omega B_0}{qN_s}\, \chi^{(1)}(\omega)\,\chi^{(1)}(\omega)
\end{align}
where $+$ and $-$ correspond to left and right circular polarizations respectively from the viewpoint of the receiver and $\chi^{(1)}\equiv N_s \alpha^{(1)}$. In the limit $\gamma \ll \omega$ (which is true for most materials at optical frequencies), these susceptibilities can be written as
\begin{align}
\chi_{\pm} &= \chi^{(1)} \pm  \omega_L \frac{\part\chi^{(1)}}{\part\omega}\; . \label{eq:chipm}
\end{align}
where $\omega_L \equiv \xi_2 qB_0/(2m)$. Comparing this to a Taylor expansion of $\chi^{(1)}$ reveals
\begin{align}
\chi_{\pm}(\omega) &\approx \chi^{(1)}\left(\omega \pm \omega_L\right)
\end{align}
assuming $\omega_L$ is sufficiently small. In normal matter $m_e \ll m_n$ so $\xi_2 \rightarrow 1$ and the applied magnetic field shifts the resonance by $\pm q B_0 / (2m_e)$ which was the result Lorentz first obtained in 1896 \cite{Lorentz1897}. Lorentz shared his theory with P.\ Zeeman which was used to calculate the charge to mass ratio of the oscillating charged particles in matter months before J.\ J.\ Thomson's discovery of the ``corpuscle'' and calculation of its charge to mass ratio \cite{Zeeman1896b,Thomson1897}. In the same year, J.\ Larmor obtained the same theoretical result as Lorentz using a classical model where the electron orbits a nucleus \cite{Larmor1897}. Furthermore, this shift is surprisingly the same as one predicted by a quantum mechanical treatment of a spin-less electron in an $l=1$ state \cite{Griffiths2005}. Using the known electron charge and mass, the linear Zeeman shift predicted by the LOM is approximately $1$ MHz/G. Of course, the LOM cannot explain the ``anomalous'' Zeeman effect because spin is fundamental to the latter but neglected in the former.

Shortly after the discovery of the Zeeman effect it was realized that the Faraday and linear Zeeman effects were both manifestations of the same physical interaction which lead H.\ Becquerel to derive his expression for the Verdet constant \cite{Becquerel1897}. The corresponding refractive indices for Eq.\ \eqref{eq:chipm} are:
\begin{align}
n_{\pm} &= n_0 \pm \omega_L \frac{\part n_0}{\part\omega}\; , \label{eq:npm}
\end{align}
where $n_0^2 \equiv 1+ \chi^{(1)}$. The specific rotatory power for Faraday rotation is $\rho \equiv \pi (n_+ - n_-)/\lambda$ where $\lambda$ is the vacuum wavelength \cite{Yariv1984}. Using Eq.\ \eqref{eq:npm} this can be rewritten as $\rho = \mathcal{V}B_0$ where
\begin{align}
\mathcal{V} &= -\frac{\xi_2q\lambda}{2mc} \frac{\part n_0}{\part\lambda} \; ,
\end{align}
is the Verdet constant. This impressively accurate expression (with $\xi_2\rightarrow 1$) was derived by Becquerel one year after Zeeman's discovery of the Zeeman effect \cite{Becquerel1897}. Ultimately the connection between the Faraday and linear Zeeman effects exemplifies the relationship between the real and imaginary parts of the refractive index or susceptibility illustrated by the Kramers-Kronig relations. The shift of an absorption resonance must be accompanied by a change in phase velocity off-resonance \cite{Fitzgerald1898}.

In atomic Positronium, $m_e = m_n$ so $\xi_2 \rightarrow 0$ and the LOM correctly predicts that atomic Positronium should have zero linear ``normal'' Zeeman shift \cite{Landau1971}.

\section{\label{sec:SP}Stark and Pockels effects}

Exploring the connection between the linear Stark and Pockels effects can be done by following a procedure similar to that used in \ref{sec:fzdetails}, assuming the optical electric field is weak enough, and using Eq.\ \eqref{eq:P1}, \eqref{eq:P2}, and \eqref{eq:chi2}. However, this approach depends critically on both the symmetry of the medium and the orientation of the applied quasi-static electric field. To keep the discussion as general as possible these details will be ignored. Instead, by comparing the first and last terms in Eq.\ \eqref{eq:P2} (and using Eqs.\ \eqref{eq:P1}, \eqref{eq:chi2}, and \eqref{eq:chipeb} and the results of \ref{sec:fzdetails}) one may estimate the magnitude of the resonance shift as
\begin{align}
\omega_S &\equiv \frac{q k^{(2)} E_0}{m^2\omega_0^3} \; .
\end{align}
Using the typical estimation of $k^{(2)} = m\omega_0^2/d$ where $d$ is the lattice constant (see Ref.\ \cite{Boyd2008}) the linear Stark shift is approximately
\begin{align}
\omega_S &\approx \frac{qE_0 \lambda_0}{2\pi m c d} \label{eq:Starky}
\end{align}
where $\lambda_0$ is the vacuum wavelength corresponding to the resonance. This expression agrees well with real experimental values. For example, using a ``lattice'' constant of the Bohr radius Eq.\ \eqref{eq:Starky} predicts a linear Stark shift of $36$ GHz/MV/m for the Lyman-$\alpha$ line in atomic Hydrogen whereas the real value is $40$ GHz/Mv/m \cite{Griffiths2005}. As another example, the predicted linear Stark shift for Nitrogen-vacancy defect centers in diamond is $26$ GHz/MV/m while the measured value is approximately $6.3$ GHz/MV/m \cite{Tamarat2006}. 

Interestingly atomic Hydrogen exhibits a linear Stark effect despite the Coulomb force being spherically symmetric which seems to contradict the attribution of the Stark effect to an anharmonic restoring force. However, it is well known that the first excited states of atomic hydrogen that experience linear Stark shifts also possess permanent electric dipole moments even in the absence of an applied external field \cite{Landau1965}. This implies a lack of inversion symmetry, which is ultimately due to the angular momentum of each state. For the LOM this implies the restoring forces representing these transitions must lack inversion symmetry. In general, time-reversal symmetry guarantees  that  any  non-degenerate  state  cannot  have  a  permanent  electric  dipole moment hence no linear Stark shift either \cite{Klemperer1993}.


As shown above, the linear Stark effect appears to be described both qualitatively and quantitatively by an anharmonic restoring force (and hence $\boldsymbol{\chi}^{(2)}$). Indeed the only form of constitutive relation that can describe the linear Stark effect is the last term in Eq.\ \eqref{eq:P2}. Since the Pockels effect is well described by the off-resonant part of the same susceptibility this supports the claim that the Pockels and linear Stark effects are simply different manifestations of the same physical interaction and that they are related through the Kramers-Kronig relations.

\section{\label{sec:photmom}Photon momentum in matter}

The momentum transferred per photon in matter can be determined by dividing the time-averaged rate of momentum transferred by a plane wave by the time-averaged rate at which photons are absorbed. Although the LOM is purely classical, photons can be realized by positing the quantization of the field energy in units of $\hbar \omega_0$ per photon. The time-averaged mechanical power absorbed from arbitrary fields by a single oscillator is then
\begin{align}
\left\langle \mathcal{P}_m \right\rangle &= \left\langle \mathcal{N} \right\rangle \hbar \omega_0 \; , \label{eq:photpower}
\end{align}
where $\langle \mathcal{N} \rangle$ is the average number of photons absorbed per unit time and the angled brackets denote time-averaging. On the other hand, the time-averaged rate of mechanical work done on each particle by the fields via the Lorentz force may be written as
\begin{align}
\left\langle\mathcal{P}_n \right\rangle &= \left\langle \dot{\Br}_n \cdot \BF_{Ln} \right\rangle  & \left\langle\mathcal{P}_e \right\rangle &= \left\langle \dot{\Br}_e \cdot \BF_{Le} \right\rangle \; .
\end{align}
The velocities of each particle may be expressed in the center-of-mass coordinate system using Eqs. $\eqref{eq:defp}$ and their perturbation solutions in Eqs.\ \eqref{eq:v1}-\eqref{eq:p2}. To first order 
\begin{align}
\dot{\Br}_n &= \frac{m_e}{qM}\dot{\Bp}^{(1)} & \dot{\Br}_e &= -\frac{m_n}{qM}\dot{\Bp}^{(1)} \; .
\end{align}
Similarly, the Lorentz forces can be Taylor expanded about the center-of-mass in the same manner as Sec.\ \ref{sec:eomsC} and the perturbation solutions for $\Bp$ and $\BR$ inserted. Up to second-order
\begin{align}
\BF_{Ln}^{(1,2)} =& q\BE + \left(\frac{m_n}{M} - \xi_2\right)\bigg[(\Bp^{(1)}\cdot\grad)\BE + \dot{\Bp}^{(1)}\times\BB \bigg] \nonumber \\
&- \Bk^{(2)} : \Bp^{(1)} \Bp^{(1)} \label{eq:Lorentzforce1} \\
\BF_{Le}^{(1,2)} =& -q\BE + \left(\frac{m_e}{M} + \xi_2\right)\bigg[(\Bp^{(1)}\cdot\grad)\BE + \dot{\Bp}^{(1)}\times\BB \bigg] \nonumber \\
&+ \Bk^{(2)} : \Bp^{(1)} \Bp^{(1)} \; . \label{eq:Lorentzforce2}
\end{align}
So the leading order powers delivered to each particle are
\begin{align}
\left\langle\mathcal{P}_n \right\rangle &= \frac{m_e}{M} \left\langle \dot{\Bp}^{(1)} \cdot \BE \right\rangle  & \left\langle\mathcal{P}_e \right\rangle &= \frac{m_n}{M} \left\langle \dot{\Bp}^{(1)} \cdot \BE \right\rangle \; .
\end{align}
For ordinary matter $m_e \ll m_n$ which implies that all of the energy delivered by light goes to the electron and not the nucleus. This limit will be taken throughout the remainder of this section for simplicity.

Using the solution to Eq.\ \eqref{eq:p1}, the total power absorbed by an isotropic oscillator from a plane wave with angular frequency $\omega$ and Jones vector $\boldsymbol{\mathcal{E}}$ is
\begin{align}
\left\langle\mathcal{P}_m \right\rangle &= \frac{\varepsilon_0\omega\alpha''}{2} \,|\boldsymbol{\mathcal{E}}|^2\; ,
\end{align}
where $\alpha'' \equiv \Im{[\alpha(\omega)]}$. Comparing this result to Eq. $\eqref{eq:photpower}$ reveals that
\begin{align}
\left\langle \mathcal{N} \right\rangle &= \frac{\varepsilon_0\alpha''}{2\hbar} \, |\boldsymbol{\mathcal{E}}|^2 \; . \label{eq:photabs}
\end{align}
As expected $\left\langle \mathcal{N} \right\rangle$ is proportional to the imaginary part of $\alpha^{(1)}$ (and hence the absorption coefficient) and to the square of the electric field amplitude (and hence the optical power).

By Newton's second law, the Lorentz forces $\BF_{Ln}$ and $\BF_{Le}$ (see Eqs. $\eqref{eq:Lorentzforce1}$ and $\eqref{eq:Lorentzforce2}$) are the rates at which momentum is being transferred from the fields to the nucleus and electron respectively. For centrosymmetric, ordinary matter where $m_e \ll m_n$ the time-averaged forces produced by arbitrary fields are
\begin{align}
\left\langle\BF_{Ln}^{(1,2)}\right\rangle &= \boldsymbol{0} \label{eq:forcedensn} \\
\left\langle\BF_{Le}^{(1,2)}\right\rangle &= \left\langle(\Bp^{(1)}\cdot\grad)\BE\right\rangle + \left\langle\dot{\Bp}^{(1)}\times\BB\right\rangle \; . \label{eq:forcedens}
\end{align}
Again all of the momentum is delivered to the electron because only it is accelerated by the fields. The restoring force then transfers momentum to the nucleus. For the same plane wave considered earlier with complex wave vector $\Bk = (k'+ik'')\hat{\Bk}$,
\begin{align}
\left\langle\BF_{Le}^{(1,2)}\right\rangle &= \frac{\varepsilon_0}{2} \left(\alpha''k' - \alpha'k''\right)|\boldsymbol{\mathcal{E}}|^2\,\hat{\Bk}\; . \label{eq:forcedense2}
\end{align}
Suppose that the frequency of the plane wave is resonant with the oscillator as is the case for common PDE experiments where resonant light excites free carriers \cite{Gibson1970,Danishevski1970,Gibson1980}. Then, $\alpha'=0$ and dividing Eq. $\eqref{eq:forcedense2}$ by Eq.\ $\eqref{eq:photabs}$ reveals that each resonant electron receives a momentum of
\begin{align}
\left\langle\boldsymbol{\Upsilon}\right\rangle_M &= \frac{n'\hbar\omega}{c} \, \hat{\Bk} \; , \label{eq:minkow}
\end{align}
per photon where $n^2\equiv 1 + \chi^{(1)}(\omega)$ is the square of the refractive index at $\omega$ and $k=n\omega/c$ was assumed. This is precisely the Minkowski value.

However, the total momentum of the photon (which is determined by the refractive index) is still unknown. To find it, a similar approach can be taken but the entire material (treated as a collection of multiple oscillator species) must be considered instead of just a single oscillator. Equations that are the volume density equivalents of Eq. $\eqref{eq:forcedense2}$ and Eq.\ $\eqref{eq:photabs}$ can then be obtained (the left-hand-sides are now a force density and a number of photons absorbed per unit volume while on the right-hand-side $\alpha\rightarrow\chi$). Taking their ratio reveals the total momentum per photon is
\begin{align}
\left\langle\boldsymbol{\Upsilon}\right\rangle_{AM} &= \frac{\hbar\omega}{c} \, \frac{n' + n'^{-1}}{2} \, \hat{\Bk} \; . \label{eq:totphotmom}
\end{align}
where $n''^2 \ll 1$ was assumed. Eq.\ \eqref{eq:totphotmom} implies that when photons are inside a material their total momentum is the average of the Abraham and Minkowski values which agrees with other derivations \cite{Mansuripur2004,Loudon2005}.

\section{\label{sec:photangmom}Photon angular momentum in matter}

The angular momentum of a photon in matter can be determined following a similar procedure to Appx. \ref{sec:photmom} but tracking torque instead of force. To leading order the time-averaged torque in the center of mass frame from the Lorentz force on each particle is
\begin{align}
\boldsymbol{\tau}_n&\approx \frac{m_e}{M} \; \Bp^{(1)} \times \BE & \boldsymbol{\tau}_e &\approx \frac{m_n}{M} \; \Bp^{(1)} \times \BE \; .
\end{align}
In the limit of $m_e \ll m_m$, then 
\begin{align}
\boldsymbol{\tau}_n&= \boldsymbol{0} & \boldsymbol{\tau}_e &= \Bp \times \BE \; ,
\end{align}
and all of the angular momentum is delivered to the electron.

A plane wave with Jones vector $\boldsymbol{\mathcal{E}}$ delivers a time-averaged torque to the electron of an isotropic oscillator of
\begin{align}
\left\langle\boldsymbol{\tau}_e\right\rangle &= \frac{i\varepsilon_0\alpha''}{2}  \,\boldsymbol{\mathcal{E}} \times \boldsymbol{\mathcal{E}}^* \; . \label{eq:torque}
\end{align}

Dividing Eq. $\eqref{eq:photabs}$ by Eq.\ $\eqref{eq:torque}$ we find the average angular momentum delivered per absorbed photon is
\begin{align}
\left\langle \boldsymbol{\mathcal{L}}\right\rangle &= i\hbar\, \hat{\boldsymbol{\mathcal{E}}}\times \hat{\boldsymbol{\mathcal{E}}}^* \; . \label{eq:photangmom}
\end{align}
For linearly polarized light, $\left\langle \boldsymbol{\mathcal{L}}\right\rangle = \boldsymbol{0}$ while for left (right) circularly polarized light $\left\langle \boldsymbol{\mathcal{L}}\right\rangle$ is $\hbar \hat{\Bk}$ $(-\hbar \hat{\Bk})$. Note that the transferred angular momentum does not depend on the photon energy nor the refractive index like the transferred linear momentum does.

The total angular momentum of a photon in matter can be calculated by considering a material composed of many oscillators rather than a single oscillator. One finds the photon's total angular momentum is also given by Eq.\ \eqref{eq:photangmom}. This implies the angular momentum is only transferred to the electrons responsible for absorption (this contrasts with the PDE).

\bibliographystyle{unsrt}
\bibliography{LOMbib}

\end{document}